\documentclass[journal,10pt,comsoc]{IEEEtran}
\usepackage{amssymb}
\usepackage{stfloats}
\usepackage{amsmath}
\usepackage{amsfonts}
\usepackage{graphicx}
\usepackage{algorithm}
\usepackage{algorithmic}
\usepackage{cite}
\usepackage{epstopdf}
\usepackage{color}
\usepackage{multirow}
\usepackage{enumerate}
\newtheorem{theorem}{\textbf{Theorem}}

\newtheorem{lemma}{\textbf{Lemma}}

\newtheorem{proposition}[theorem]{\textbf{Proposition}}
\newtheorem{remark}{\textbf{Remark}}

\begin{document}
\title{Design and Optimization of VoD schemes with Client Caching in Wireless Multicast Networks}
\author{Hao Feng,~\IEEEmembership{Student Member,~IEEE},  Zhiyong Chen,~\IEEEmembership{Member,~IEEE,} and Hui Liu,~\IEEEmembership{Fellow,~IEEE}\\
\thanks{This work has been partly presented in IEEE GLOBECOM $2016$ \cite{conference}.  H. Feng, Z. Chen (\textit{corresponding author}), and H. Liu are with the Department of Electronic Engineering, Shanghai Jiao Tong University, Shanghai,  China.  Emails: \tt \{fenghao, zhiyongchen, huiliu\}@sjtu.edu.cn}}
\maketitle
\begin{abstract}
Due to the explosive growth in multimedia traffic, the scalability of video-on-demand (VoD) services is become increasingly important. By exploiting the potential cache ability at the client side, the performance of VoD  multicast delivery can be improved through video segment pre-caching. In this paper, we address the performance limits of client caching enabled VoD schemes in wireless multicast networks with asynchronous requests.  Both  reactive and proactive systems are investigated. Specifically, for the reactive system where videos are transmitted on demand, we propose a joint cache allocation and multicast delivery scheme to minimize the average bandwidth consumption under the zero-delay constraint. For the proactive system where videos are periodically broadcasted, a joint design of  the cache-bandwidth allocation algorithm and the delivery mechanism is developed to minimize the average  waiting time under the total bandwidth constraint. In addition to the full access pattern where clients view videos in their entirety, we further consider the access patterns with random endpoints, fixed-size intervals and downloading demand, respectively. The impacts of different access patterns on the resource-allocation algorithm and the delivery mechanism are elaborated. Simulation results validate the accuracy of the analytical results and also provide useful insights in designing  VoD networks with client caching.
\end{abstract}
\section{Introduction}
The rapid proliferation of smart devices has led to an unprecedented growth in internet traffic.  According to Cisco's most recent report\cite{cisco}, the traffic data of video-on-demand (VoD) services is forecast to grow at a compound annual growth rate of more than $60\%$.  However, the  traditional unicast-based delivery mechanism, where a server responds to each client individually, is unlikely to keep pace with the ever-increasing traffic demand. On the other hand,  the traffic demand for videos, although massive and ever-increasing, is highly redundant, i.e., the same  video is requested multiple times and a small number of videos account for a majority of all requests\cite{Zipf}.  Therefore, a promising approach is to deliver these popular videos to multiple clients via multicast.

VoD multicast delivery  has attracted significant interest  recently. In industry, apart from the broadcasting networks,  the evolved multimedia broadcast/multicast service (eMBMS) is introduced in the  long term evolution (LTE) networks\cite{embms}.  In academic, extensive studies  have been conducted on the efficient  multicast delivery for VoD services\cite{svc_ton,comulticast,vodsurvey,jsac_patch}. Among them, one aspect is to provide the reliable and efficient multicast delivery to clients with synchronous requests for the same videos, such as scalable video coding design and cooperative multicast mechanism\cite{svc_ton,comulticast}. Another important issue is to design  bandwidth efficient multicast delivery schemes to meet asynchronous requests at different times,  including batching, patching, stream merging and periodic broadcasting\cite{vodsurvey,jsac_patch},  which  is the main focus of this paper.

In addition to  the VoD multicast delivery, another important trend is that the cache capacity at client side  is  increasing rapidly and should be effectively exploited\cite{threecolors,kongtao}. Therefore, client storage could not only be  used as a traditional short-term memory which  temporarily buffers ongoing desired video segments
at the client request times\cite{re01batch1,re12BUPT,re01hsm}, but also  serve as  a  long-term memory to pre-cache initial popular video segments  ahead of client request times\cite{Niulab}.  In this case, {the bandwidth consumptions at the server and the network sides are greatly reduced, and also the average client waiting time can  be highly saved\cite{Melbourne,pro99zero,pro06GEBBPC,unicastcache,codedcache1}.}  In this paper,  we will explore the  optimal combination  of client caching and multicast delivery   for improving the scalability of  VoD systems.

\subsection{Related Work}
In general,  existing VoD multicast schemes   for asynchronous requests fall into  two transmission modes\cite{Mcache}, i.e., reactive and proactive modes. Reactive mode implies that the delivery system is two-way in nature and there exists an uplink channel to report client requests. In this mode,  videos are transmitted on demand\cite{re01batch1,re12BUPT,re01hsm,Niulab}. Proactive mode means that the delivery system is only one-way and has no uplink channel to report client demands. In this mode, videos are periodically broadcasted with predefined carouse periods\cite{pro1997skyscraper,pro98fastdata,pro97harmonic}.  

For the reactive  system, various delivery schemes have been proposed in past decades, including but not limited to batching, patching and merging\cite{re01batch1,re12BUPT,re01hsm}. In batching, requests for the same video are delayed for a certain time so that more requests can be served concurrently within one multicast stream\cite{re01batch1}. In patching, a client joins a desired ongoing multicast stream, and a unicast/multicast stream is established to patch the missing part\cite{re12BUPT}. In merging,  a client  could join several ongoing multicast streams and the patching streams {of different clients} are merged into one multicast stream\cite{re01hsm}. Among these techniques, \cite{re01hsm} proves  the optimality of merging in terms of the  minimum bandwidth requirement. Subsequently, \cite{re03packetloss} extends this technique to wireless channels based on erasure codes. For multi-video delivery,    \cite{re07hybrid} and  \cite{re09TOBhybridWimax} propose hybrid transmission mechanisms where popular videos are periodically broadcasted and less popular videos are served via either grace-patching or unicast.  These studies \cite{re01batch1,re12BUPT,re01hsm,re03packetloss,re07hybrid,re09TOBhybridWimax} utilize the client storage, however, only for  temporary buffering and the potential cache capacity at the client side is not fully exploited.  By pre-caching initial video segments into the client storage,  \cite{Niulab} adopts the batching delivery mechanism and optimizes the cache allocation to minimize the energy consumption.  By buffering an ongoing stream and receiving a   multicast patching stream, \cite{Melbourne} further  proposes the prepopulation assisted batching  with multicast patching  (PAB-MP). {These studies} \cite{Niulab,Melbourne} are based on either batching or patching, and the optimal cache allocation algorithm and the corresponding delivery mechanism  for VoD services with client caching  are still unknown.  

For the proactive system, various periodic broadcasting schemes have been well studied, including skyscraper broadcasting\cite{pro1997skyscraper}, fast broadcasting\cite{pro98fastdata} and harmonic broadcasting\cite{pro97harmonic}.  In all these schemes, videos are divided into  a series of segments and  each segment is broadcasted
periodically on dedicated subchannels. Hu in \cite{pro2001GEBB}  firstly derives the theoretical lower bandwidth requirement bound for any periodic broadcasting protocols and proposes a  greedy equal bandwidth broadcasting (GEBB)  scheme that achieves the minimum fixed delay under the bandwidth constraint. Reference \cite{pro06reliable}  further applies the GEBB scheme with fountain codes to wireless systems. However, in \cite{pro2001GEBB} and \cite{pro06reliable}, the client storage is only utilized for temporary buffering  and the potential cache capacity at the client side is not effectively exploited.   By pre-caching  initial video segments at the client side, \cite{pro99zero} and \cite{pro06GEBBPC} develop zero-delay delivery schemes  based on  polyharmonic broadcasting and GEBB, respectively. {Reference} \cite{pro14BackgroundPush} investigates the cache-bandwidth allocation  and the delivery mechanism for  multi-video  delivery in digital video broadcasting (DVB) systems. However, the proposed  delivery mechanism is designed for video downloading instead of streaming, and the  cache-bandwidth allocation is  not jointly optimized.
\subsection{Motivation and Contributions}
Despite the aforementioned studies, the following fundamental questions regarding VoD services in  reactive and proactive systems with client caching remain unsolved to date. \emph{Q1: What are the optimal reactive and proactive multicast delivery mechanisms when the  cache capacity at the client side can  be exploited?}\emph{ Q2: What is the corresponding optimal resource (e.g., cache and bandwidth) allocation for multi-video delivery?} In addition, the aforementioned studies rest on the assumption of the full access pattern where clients watch the desired video from the beginning to the end. However, clients might be interested in video intervals rather than full-videos\cite{part}.  References \cite{re2002limitspatterns} and \cite{re2002fixedsize} consider  interval access patterns with random intervals and fixed-size intervals for  VoD services without client caching, respectively. To the best of our knowledge, the impacts of different access patterns on the resource allocation algorithm and the delivery mechanism  for VoD services with client caching are also unknown.

In this paper, we attempt to answer the above key questions and  provide the performance limits of VoD multicast schemes for both reactive and proactive systems where clients have certain cache capacity. Both full and interval access patterns are investigated.  Our main contributions are  as follows:

\begin{itemize}
\item
\textbf{Optimal joint cache allocation and multicast delivery scheme for the reactive system: }  In  Sec. \ref{sec:reactive},  a joint cache allocation and multicast delivery scheme is developed to minimize the average bandwidth consumption of VoD services  in the reactive system under the zero-delay constraint. We first propose a client caching enabled multicast patching (CCE-MP)  mechanism which minimizes the average bandwidth consumption given a certain cache allocation. Then  we formulate the cache allocation problem under the full access pattern  into a convex problem, which can be effectively solved  by a water-filling algorithm. {\color{black}This analysis provides a useful insight in choosing the minimum bandwidth-cache resource to meet a certain client request rate}.
\item
\textbf{Optimal joint cache-bandwidth allocation and multicast delivery scheme for the proactive system}:
In Sec. \ref{sec_proactive}, we jointly design the cache-bandwidth allocation algorithm and the multicast delivery mechanism to minimize the average client waiting time for the  proactive  system under  the total bandwidth constraint.  Firstly we propose a client caching enabled GEBB (CCE-GEBB) delivery mechanism and show its optimality in term of the minimum client waiting time given a certain cache-bandwidth allocation.  By exploring the structure of the optimal solution, we then simplify the  cache-bandwidth allocation problem under the full access pattern   to a one-dimensional search of the allocated cache size for the most popular video.
\item
\textbf{Impact of different  client access patterns}: We investigate the impacts of different access patterns on the resource allocation algorithm and the multicast delivery mechanism in both reactive and proactive systems.  In addition to the content popularity, the optimal scheme also depends on the client request rate  and the access pattern in the reactive  system.  For instance, it is optimal to cache  videos evenly for the full access pattern and the  interval access pattern with random endpoints under relatively  high  request rates. Meanwhile, caching simply the most popular videos is optimal for the full access pattern under relatively low  client request rates and the  fixed-size interval access pattern under all request rates.
\end{itemize}
\subsection{Organization}
The remainder of this paper is organized as follows. Section \ref{sec:system} introduces the system model of the VoD delivery network with client caching. Sections \ref{sec:reactive} and  \ref{sec_proactive} present the optimal resource allocation and multicast delivery schemes  in reactive and proactive  systems, respectively. Simulation results are shown in Section \ref{sec:numerical} and we conclude in Section \ref{sec:conclusion}.
\vspace{-1mm}
\section{System Model}\label{sec:system}
\begin{figure}[t]
\centering
\includegraphics[width=2.6in]{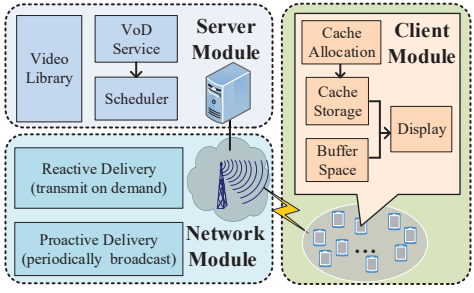}
\vspace{-1mm}
\caption{ System model for the video delivery network with client caching.}\label{fig:system_model}
\vspace{-3mm}
\end{figure}

 Fig.  \ref{fig:system_model} shows the simplified logical architecture of the VoD multicast delivery network, which includes a server module, a network module and a client module.
\subsection{Server Module}
The video server contains a  library of $M$  constant bitrate (CBR) videos $\mathcal{V}=\{V_1,\ldots,V_M\}$, and  each video is characterized by a  tuple \{\textit{length}, \textit{bitrate}, \textit{popularity}\}, where the popularity is defined as the video access probability.  Note that  we consider CBR videos in this paper for simplicity, same as \cite{unicastcache,Niulab,Melbourne}. Also, as indicated in \cite{youtube}, the videos in YouTube are encoded into CBR.  Moreover, the analysis in this paper can be extended to variable bit rate (VBR) videos since a VBR video can be regarded as a collection of different CBR video chunks and  regulated into CBR streams\cite{TON_CBR}.

Denote the length and the bitrate of $V_i$ as $L_i$ and $r_i$, respectively.  Since the main purpose of this paper is to reveal the relationship between resource allocation and the content popularity, and also the relationship between length (or bitrate) and popularity of video has not been explicitly discovered so far\cite{unicastcache},  we assume that all the videos are of equal length $L$ and bitrate $r$  for simplicity, i.e., $L_1=\ldots=L_M=L$ and $r_1=\ldots=r_M=r$. Note that the extension to the general case where videos are of different lengths and bitrates is quite straightforward. The popularity distribution vector of the videos is denoted by $\mathbf{p}=[p_1,\ldots,p_M]$, which is assumed to be known apriori (e.g., estimated via some learning
procedure\cite{zipf_training}), where $\sum\limits_{i=1}^{M}p_i=1$. In addition, we assume $p_1\geq p_2\ldots \geq p_M$, i.e., the popularity rank of $V_i$ is $i$.
\subsection{Network Module}
The transport network is aimed to reliably multicast the desired videos from the server to clients under coverage, which can be guaranteed by forward error correction (FEC) codes at application and physical layers\cite{re03packetloss,pro06reliable}. The corresponding bandwidth efficiency is denoted by $f_B$ (bps/Hz). Depending on whether clients report their demands to the server or not, the  network can be divided into the following two types:
\subsubsection{\textbf{Reactive  System}}
The transport network is \textit{reactive} delivery if  it is a two-way transmission system, and  videos are delivered in response to client requests.
\subsubsection{\textbf{Proactive  System}}
The transport network is \textit{proactive} delivery if it is one-way  and clients have no uplink channel to report their demands. In this case, videos are broadcasted periodically at different carouse  periods, e.g.,  videos with larger popularity are transmitted more frequently.  Note that the proactive delivery can be regarded as a special case of  the reactive delivery when the gathered request  information is not exploited in the reactive  system and the server simply broadcasts videos periodically.
\vspace{-2mm}
\subsection{Client Module}
\vspace{-1mm}
Each client is equipped with a cache storage and a buffer space.  The cache storage is used as a long-term memory to cache video segments ahead of client request times{\footnote{\color{black}In addition to client caching, caching within the network (e.g., at the proxy or base stations (BSs)) is also a promising approach. In this paper, we only consider client caching, while BS caching is out of scope here. Note that although BSs with large cache storage can help to save the wired band and the transmission latency from the server to the BSs,  it does not actually reduce the wireless link traffic from the BS to the client side, where the wireless  capacity is the bottleneck. }}, while the buffer space is  a short-term memory which temporarily buffers ongoing desired uncached data
at the client request times  and the buffered data would be released right after consumed.  We assume  that all clients have the same  cache size $C$, where $C\!<\!ML$. The cache allocation of the videos is denoted by $\mathbf{l}=[l_1,\ldots, l_M]$, where $l_i$ is the allocated cache size for storing  the $i$-th video.  We then have the storage constraint $\sum\limits_{i=1}^{M}{l_i}\!\leq \!C$. 

The client request events are modeled as a Poisson process with parameter $\lambda$, and clients request videos according to the video popularity distribution $\mathbf{p}$. Since Poisson processes remain Poisson
processes under merging and splitting, the request arrivals for $V_i$ also follow a Poisson process with  $\lambda_i$, where $\lambda_i=p_i\lambda$. When a client demands a certain video, it first checks whether the desired video has been cached, and if so,  the client would display the cached part locally while  buffering the ongoing uncached part. For the uncached part,  the client  in the reactive system sends the request  to the server and the server will  transmit the desired part at a suitable time,   while the  client in the proactive system needs to wait for the scheduling of the desired part in the periodic broadcasting.

Moreover, different  access patterns should be considered since clients might only view a part of a video, e.g., they lose interest  and stop watching before the end of the video. In this paper, we consider the following two typical  access patterns:
\begin{itemize}
\item \textbf{Full Access Pattern}:  The client  views the desired video entirely, i.e., from the beginning to the end.
\item \textbf{Interval Access Pattern}: Client requests are for video intervals rather than full-videos. e.g., they  watch a video from the same beginning but end the watching   at random endpoints (i.e., the access pattern with random endpoints\cite{re2002limitspatterns}), or they  watch a video only for fixed-size intervals from random beginnings (i.e., the fixed-size access pattern\cite{re2002fixedsize}). In addition, clients may  watch interested video clips until the video is fully saved (i,e., the downloading-demand access pattern\cite{pro14BackgroundPush}). In this paper, we consider the interval access patterns with random endpoints and fixed-size intervals for  the reactive system, while the access patterns with random  endpoints and  downloading demand are investigated for  the proactive  system.
\end{itemize}
\vspace{-2mm}
\section{VoD Delivery in the Reactive  System}\label{sec:reactive}
\vspace{-1mm}
In this section, we devise a joint  cache allocation and multicast delivery scheme  to minimize the  average bandwidth consumption for VoD services under the zero-delay constraint. {\color{black}Note that  lower bandwidth consumption implies  more clients can be supported and thus highly scalable}. Both   full and interval access patterns are investigated.

Since no multicast opportunity exists among different video demands,  the bandwidth consumption of  each video can be acquired individually.  The bandwidth consumption minimization  problem can be  written in the following general form:
\begin{align}
    \min_{\mathbf{l},S_{rd}}& ~~\sum\limits_{i=1}^{M}{b}_i(l_i,S_{rd})\label{pro:proactive_general}\\
    \textup{s.t.}&  ~~\sum\limits_{i=1}^{M}{l_i}\leq C, \label{pro:cache_constraint1} \\
     &~~ 0 \leq  l_i\leq L, \forall i\in\{1,\ldots,M\}\label{pro:cache_constraint2},
\end{align}
where ${b}_i(l_i,S_{rd})$ is the  average bandwidth  consumption of the $i$-th video $V_i$ with cache size $l_i$ and reactive delivery mechanism $S_{rd}$  under the zero-delay  constraint.

We first introduce the optimal multicast delivery mechanism which  minimizes the  average bandwidth consumption  of $V_i$ with cache size $l_i$ and request rate $\lambda_i$ under any access pattern.
\begin{proposition}\label{pro:CCE-MP}
The optimal  multicast delivery mechanism, referred as the  client caching enabled multicast patching (CCE-MP) mechanism,  consists of the following two operations:  a) the server multicasts every desired uncached part at the latest deadline (i.e., at the time of display); b)  each client starts buffering the desired uncached data from any ongoing multicast stream right after the client request time.
\end{proposition}
\begin{figure}
\centering
\includegraphics[width=3.2in]{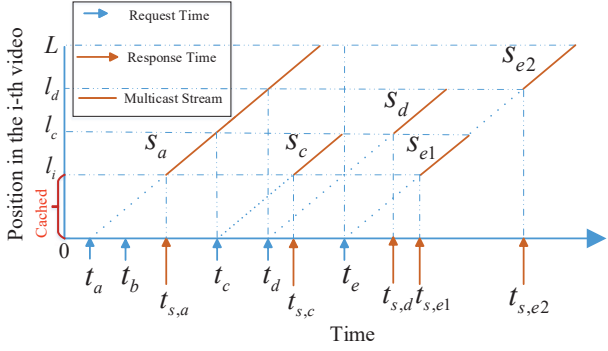}
\vspace{-3mm}
\caption{Example of CCE-MP under the full access pattern.}\label{fig:OPC}
\vspace{-4mm}
\end{figure}
\begin{IEEEproof}
See Appendix A.
\end{IEEEproof}

An example illustrating the basic operations of CCE-MP under the full access pattern is provided in Fig. \ref{fig:OPC}. In the figure, clients A through E intend to view $V_i$ entirely  at different times, i.e., $t_a$ through $t_e$. The beginning part with length $l_i$ has been pre-cached at the client side. For the first request at  time $t_a$ by client A, the server does not respond immediately since client  A could enjoy the beginning part locally due to prefix cache. In  this case,  the latest time to schedule  a patching stream $s_a$ with transmission rate $r$  should be $t_{s,a}=t_a+{l_i}/{r}$, right  after client A finishes local display, where $r$ is also the slope of orange solid lines in Fig. \ref{fig:OPC}. Meanwhile, client B whose request arrival time is within $[t_a, t_{s,a}]$ is also  satisfied by the same multicast stream $s_a$. When C clicks on the video at time $t_c$,  it immediately buffers the remaining part of the ongoing  stream $s_a$ from video position $l_c=l_i+r(t_c-t_{s,a})$. Then the server  schedules another patching stream $s_c$ from video position $l_i$ to position $l_c$ at time $t_{s,c}$, where  $t_{s,c}=t_c+{l_i}/{r}$ is the latest deadline for client C to receive that part.  Similarly, the server schedules  $s_d$, $s_{e1}$ and  $s_{e1}$ in response to clients D  and E.

\begin{figure}
\centering
\includegraphics[width=2.6in]{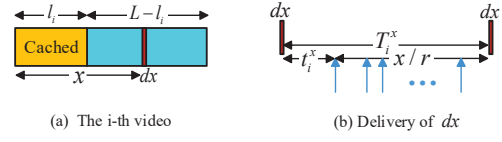}
\vspace{-1mm}
\caption{Illustration of the derivation of the average bandwidth consumption.}\label{fig:lowerbound}
\vspace{-3mm}
\end{figure}
\subsection{Full Access Pattern}
For this pattern, we first derive the average bandwidth consumption of the CCE-MP  mechanism given a certain cache allocation, and then obtain the  optimal cache allocation.
 \subsubsection{\textbf{Average Bandwidth Consumption of CCE-MP}}
 Similar to the optimal merging without client  caching in \cite{re01hsm}, the  average bandwidth consumption of CCE-MP can be  derived by splitting a video into arbitrary small portions and obtaining the average bandwidth consumption  of each portion individually.
 As illustrated in Fig. \ref{fig:lowerbound}, we  take the transmission of a small portion  $dx$ at an arbitrary length offset $x$ of the $i$-th video for example, where $l_i \leq x\leq L$. Let $t_i^x$ be the  time interval between the previous transmission of $dx$ and the following first video request. Let  $T_i^x$ denote the time interval between two successive transmissions of $dx$. We then have $T_i^x=t_i^x+{x}/{r}$ according to operation (a) in \textbf{Proposition \ref{pro:CCE-MP}},  since the transmission of $dx$ is triggered by the first  request and scheduled until the display reaches  position $x$.  Meanwhile, the  following requests  can all be satisfied by the same transmission of $dx$ according to operation (b). It can be verified that the transmission of $dx$ follows  a renewal process.  Let $S(t)$  denote the total data amount  for  delivering $dx$ from time $0$ to $t$. Due to the property of the renewal process, the  average bandwidth consumption for delivering $dx$ is $\bar{b}_i^x=\frac{1}{f_B}\lim\limits_{t\rightarrow \infty}\frac{S(t)}{t}=\frac{dx}{f_BE(T_i^x)}$, where $E(T_i^x)$ denotes the expectation of $T_i^x$. Therefore,  the  average bandwidth consumption of $V_i$ can be written as  $b_{i}=\int_{l_i}^{L}\frac{dx}{f_BE(T_i^x)}$.

 Under the full access pattern, all clients  watch the video entirely. Due to the memoryless property of the exponential distribution, $t_i^x$ also follows the exponential distribution with parameter $\lambda_i$. Then we have $E(T_i^x)={1}/{\lambda_i}+{x}/{r}$, and  the  average  bandwidth consumption of $V_i$ is
\begin{equation}\label{equ:full_lowerbound}
b_{i}^{FA}=\int_{l_i}^{L}\frac{1}{f_B(\frac{1}{\lambda_i}+\frac{x}{r})}dx=\frac{r}{f_B}\ln\left(\frac{L-l_i}{l_i+\frac{r}{\lambda_i}}+1\right).
\end{equation}
\begin{remark}
When $l_i=0$, we have $b_{i}^{FA}=\frac{r}{f_B}\ln(\frac{L}{r/\lambda_i}+1)$. In this case, CCE-MP reduces to the optimal merging without client caching in \cite{re01hsm}. Compared to $l_i=0$, (\ref{equ:full_lowerbound}) indicates the following two benefits of client caching: a) \textit{local cache gain} incurred by the  pre-cached part; b) \textit{multicast gain} by allowing the server to delay the delivery due to local cache  and serve a batch of requests via a single multicast stream.
\end{remark}
\begin{remark}\label{mark:retopro}
When $\lambda_i\gg\frac{r}{l_i}$, we have $E(T_i^x)=\frac{x}{r}$ and $b_i^{FA}=\frac{r}{f_B}\ln(\frac{L}{l_i})$. In this case, CCE-MP reduces to the proactive delivery mechanism which periodically broadcasts the video, i.e., portion $dx$ at offset $x$  is broadcasted every $\frac{x}{r}$ time units, where $l_i \leq x\leq L$. Therefore, $\frac{r}{f_B}\ln(\frac{L}{l_i})$ is also the minimum average bandwidth consumption of $V_i$ with cache size $l_i$ under the zero-delay constraint in the proactive  system.
\end{remark}

{\color{black}
Let $b_{i,batch}$ denote the average bandwidth  consumption of $V_i$ with cache size $l_i$ under the batching method. In  the batching method\cite{Niulab}, multiple  client requests for the same video that arrive within a batching window (i.e., the local displaying period due to prefix cache) are grouped and served via a single multicast transmission. According to \cite{Niulab}, we have
\begin{equation}
b_{i,batch}=\frac{r}{f_B}\frac{L-l_i}{l_i+\frac{r}{\lambda_i}},
\end{equation}
decreasing with the increase of  cache size $l_i$. When $l_i=0$, we have $b_{i,batch}=\lambda_{i}{L}/f_B$ and the batching method reduces to serving each client request via unicasting. We have the following lemma.
\begin{lemma}\label{lemma:gap}
  Compared to the batching method, the bandwidth saving of  CCE-MP becomes smaller with larger cache  size, i.e., $\frac{b_{i,batch}-b_{i}^{FA}}{b_{i,batch}}$ decreases with  increasing cache size $l_i$.
\end{lemma}
\begin{IEEEproof}
See Appendix B.
\end{IEEEproof}
}
\subsubsection{\textbf{Cache Allocation}}  The cache allocation problem is
\begin{align}
    \min_{\mathbf{l}}& ~~\sum\limits_{i=1}^{M}\frac{r}{f_B}\ln\left(\frac{L-l_i}{l_i+\frac{r}{\lambda_i}}+1\right)\label{pro:cache_fa}\\
    \textup{s.t.}&  ~~(\ref{pro:cache_constraint1}), (\ref{pro:cache_constraint2}) \nonumber.
\end{align}
\begin{figure}[t]
\centering
\includegraphics[width=2.3in]{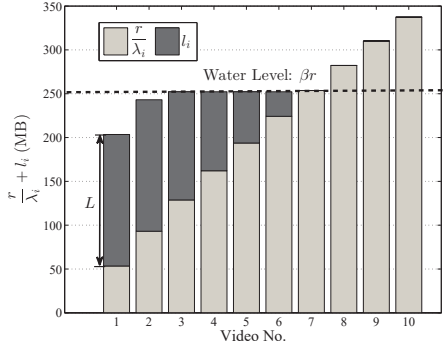}
\vspace{-1mm}
\caption{The optimal cache allocation under the full access pattern with  $M=10$, $C=0.4ML$ and $\lambda=1$ min$^{-1}$, and the other  settings are the default values of Table \ref{table:setting} in Sec. \ref{sec:numerical}.}
\label{fig:waterfilling}
\vspace{-3mm}
\end{figure}
\begin{lemma}\label{lemma:waterfill}
(\textit{Water-filling Algorithm})
The optimal cache allocation is
\begin{equation}\label{equ:waterfilling}
{\color{black}l_i=\min\left((\beta{r}-\frac{r}{\lambda_i})^+,L\right)},  \quad\text{for } i\in\{1,2,\ldots,M\},
\end{equation}
where $x^+=\max{(x,0)}$ and $\beta$  can be effectively solved by the bisection method under the storage constraint.
\end{lemma}
\begin{IEEEproof}
See Appendix C.
\end{IEEEproof}

As illustrated in Fig. \ref{fig:waterfilling}, the water-filling algorithm  allocates larger cache sizes to videos with larger popularity.  For instance, the videos with larger popularity below the water level are cached to reach either the  entire video length $L$ {\color{black}(i.e., $V_{1}$ and $V_{2}$) } or the water level {\color{black}(i.e., $V_{3}$ to $V_{6}$)}.  Meanwhile, the videos with smaller popularity above the water level, {\color{black}i.e., $V_{7}$ to $V_{10}$},  have no cache  allocated.

\subsection{Interval Access Pattern with Random Endpoints}
Here we consider the interval access pattern  where clients watch a video from the same beginning to random endpoints. For simplicity, the endpoints are uniformly distributed.
\subsubsection{\textbf{{\color{black}Average Bandwidth Consumption of CCE-MP}}}Due to the uniform distribution of the endpoints, the probability that a client finishes watching the video before portion $dx$ at offset $x$ is ${x}/{L}$.  Hence the client request rate for  $dx$ of $V_i$ is also a Poisson process with parameter $\frac{(L-x)\lambda_i}{L}$,  yielding
$E(T_i^x)=\frac{L}{(L-x)\lambda_i}+\frac{x}{r}$.
Let  $\eta_i=\sqrt{\frac{Lr}{\lambda_i}+\frac{L^2}{4}}$, the minimum average bandwidth consumption of $V_i$ with cache size $l_i$ is
\begin{align}
b_{i}^{RE}\!=\!&\frac{1}{f_B}\int_{l_i}^{L}\frac{1}{\frac{L}{(L-x)\lambda_i}+\frac{x}{r}}dx\nonumber\\
=&\small{\left(\frac{r}{2f_B}\!+\!\frac{Lr}{4\eta_if_B}\right)\!\left(\ln\left(\frac{L}{2}\!+\!\eta_i\right)\!-\!\ln\left(l_i\!+\!\eta_i-\frac{L}{2}\right)\right)}+\nonumber\\
&\!\small{\left(\!\frac{r}{2f_B}\!-\!\frac{Lr}{4\eta_if_B}\!\right)\!\left(\ln\left(\eta_i\!-\!\frac{L}{2}\!\right)\!-\!\ln\left(\eta_i\!+\!\frac{L}{2}\!-\!l_i\!\right)\!\right)}.
\end{align}
\begin{remark}
When $\lambda_i\gg\frac{r}{l_i}$, we have $\eta_i=\frac{L}{2}$ and  $b_i^{RE}=\frac{r}{f_B}\ln(\frac{L}{l_i})=b_i^{FA}$. In this case, CCE-MP also reduces to the proactive delivery mechanism which broadcasts the video periodically, and the bandwidth consumption is the same as that of the full access pattern.
\end{remark}
\subsubsection{\textbf{Cache Allocation}}Since {\small{$\frac{\partial^2{b_{i}^{RE}}}{\partial{{l_i}^2}}\!=\!\frac{r(l_i-L)^2{\lambda_i}^2+r^2L\lambda_i}{f_B(rL+Ll_i\lambda_i-{l_i}^2\lambda_i)^2}\!\geq\!0$}}, the cache allocation problem  under this  pattern is also convex, yielding the following lemma.
\begin{lemma}
The optimal cache allocation under the interval access pattern with uniformly distributed endpoints is
\vspace{-2mm}
\begin{equation}
{\color{black}l_i=\min\left(\left(\frac{\beta{r}+L}{2}-\sqrt{\frac{\left(\beta{r}-L\right)^2}{4}+\frac{Lr}{\lambda_i}}\right)^+,L\right)}
\end{equation}
 for $ i\in\{1,\ldots,M\}$, where $\beta$ can be effectively  obtained by the bisection  method.
\end{lemma}
\begin{IEEEproof}
The proof is similar to \textbf{Lemma \ref{lemma:waterfill}}.
\end{IEEEproof}
\subsection{Interval Access Pattern with Fixed-size Intervals}
In  addition to  the interval access pattern with random endpoints, we also consider the fixed-size interval access pattern with random beginnings  proposed in \cite{re2002fixedsize}\footnote{Since the general case with random size intervals  is too complex to be analyzed\cite{re2002fixedsize}, we select the fixed-size interval access pattern proposed in \cite{re2002fixedsize} to reveal the cache allocation for the access pattern with random start points.}, i.e., each request is for a segment of duration $D$ starting from a random point, and videos are cyclic, which means  access may proceed past the end of a video by cycling to the beginning of it.
\subsubsection{\textbf{{\color{black}Average Bandwidth Consumption of CCE-MP}}}
The average bandwidth consumption  under the fixed-size interval access pattern is derived based on the following proposition.
\begin{proposition}
{(Ref. \cite{re2002fixedsize}, Sec. 3.1)} The mean interval for delivering $dx$ of the $i$-th video is
\begin{equation}\label{equ:etix}
 E[T_i^x]\!=\!\sqrt{\frac{\pi{L}}{\!2r\lambda_i}}\!\mathrm{erf}\!\left(D\sqrt{\frac{r\lambda_i}{2L}}\right)\!+\!\frac{L}{Dr\lambda_i}\exp{\left(\!-\frac{D^2r\lambda_i}{2L}\right)},
\end{equation}
where $\mathrm{erf}(t)=\frac{2}{\sqrt{\pi}}\int_{0}^{y}\exp(-y^2)dy$ denotes the error function.
\end{proposition}

Note that $E[T_i^x]$ is irrelevant to video position $x$ and we can drop the upper index of $T_i^x$ as  $T_i$. This is due to the fact that all parts of the video are of equal importance for the fixed-size interval access with cyclic display.  We then  have
\begin{equation}
b_i^{FS}=\int_{l_i}^{L}\frac{dx}{f_BE(T_i)}=\frac{L-l_i}{f_BE(T_i)}.
\end{equation}
 \subsubsection{\textbf{Cache Allocation}}
The cache allocation problem is
\begin{align}\label{prop:fixedsize}
    \min_{\mathbf{l}}& ~~\sum\limits_{i=1}^{M}\frac{L-l_i}{f_BE(T_i)}\\
  \textup{s.t.}&  ~~(\ref{pro:cache_constraint1}), (\ref{pro:cache_constraint2})\nonumber,
\end{align}
{\color{black}
and we have the following lemma.
\begin{lemma}\label{lemma:fixedsizecache}
The optimal cache allocation under the fixed-size interval access pattern  is
 \begin{equation}\label{equ:cacheallocation}
 l_i=\begin{cases}L \quad &\text{if } 1 \leq i \leq k-1\\
 C\bmod L \quad &\text{if }  i= k\\
 0 \quad &\text{if } k+1 \leq i \leq M
 \end{cases},
\end{equation}
where $k=\lfloor{C}/{L}\rfloor+1$.
\end{lemma}
\begin{IEEEproof}
 Cache allocation problem (\ref{prop:fixedsize}) is equivalent to the problem
 \begin{align}
    \max_{\mathbf{l}}& ~~\sum\limits_{i=1}^{M}\frac{l_i}{f_BE(T_i)}\\
  \textup{s.t.}&  ~~(\ref{pro:cache_constraint1}), (\ref{pro:cache_constraint2})\nonumber,
\end{align}
 which belongs to  fractional knapsack problems, where the knapsack capacity is $C$ and  the value of  caching  the unit size of $V_i$  would be $\frac{1}{f_BE(T_i)}$. The optimal solution for the fractional knapsack problem is the greedy algorithm,  which  chooses the videos with the highest $\frac{1}{f_BE(T_i)}$ values and caches them up to the full length $L$ until the knapsack capacity $C$ is used up \cite{knapsack}\cite{pushbased}. Since $E(T_i)$ decreases with larger $\lambda_i$ and the videos are ranked in the descending order of the popularity, the  greedy algorithm then reduces to cache the most popular videos up to the full length $L$ until the cache storage capacity $C$  is used up, i.e., the optimal cache allocation is Eq. (\ref{equ:cacheallocation}). Thus the proof is  completed.
\end{IEEEproof}}
\begin{remark}
The optimal cache allocation here  is independent of  the total client request rate. Therefore, under any request rate, the optimal cache allocation algorithm for the fixed-size interval access pattern is to cache the most popular videos only, termed as \textbf{Popular-Cache}.
\end{remark}
\subsection{Extreme Case Analyses}\label{subsec:extreme}
Two special cases of client request rates are  investigated to provide further insight into the impact of different access patterns on the cache allocation algorithms.
\subsubsection{$\lambda\to 0$} When $\lambda$ is relatively small, no multicast opportunity exists even among client requests for the same video. In this case, the  server satisfies each  request via unicast.
\begin{itemize}
\item
Full access pattern: The average bandwidth consumption of $V_i$ under the unicast-based transmission is ${\lambda_i(L-l_i)}/{f_B}$.  The cache allocation problem  becomes
\begin{align}
    \min_{\mathbf{l}}& ~~\sum\limits_{i=1}^{M}\frac{{\lambda_i}(L-l_i)}{f_B}\label{pro:FA_smallrequest}\\
    \textup{s.t.}&  ~~(\ref{pro:cache_constraint1}), (\ref{pro:cache_constraint2}) \nonumber,
\end{align}
which requires maximizing $\sum_{i=1}^{M}{{\lambda_i}l_i}/{f_B}$, reducing to a fractional knapsack problem. Similar to  problem (\ref{prop:fixedsize}),  the optimal solution is Popular-Cache.
\item
Interval access pattern with random endpoints:  The {\color{black}average} bandwidth consumption of $V_i$  is $\frac{1}{f_B}\int_{l_i}^{L}\frac{(L-l_i)\lambda_i}{L}dx$ $=\frac{(l_i-L)^2\lambda_i}{2Lf_B}$, and the cache allocation problem becomes
\begin{align}
    \min_{\mathbf{l}}& ~~\sum\limits_{i=1}^{M}\frac{(l_i-L)^2\lambda_i}{2Lf_B}\\
    \textup{s.t.}&  ~~(\ref{pro:cache_constraint1}), (\ref{pro:cache_constraint2}) \nonumber,
  \end{align}
which is a convex problem. The optimal  solution is $l_i=(L-\frac{\beta}{\lambda_i})^+$ for $i\in\{1,\ldots,M\}$ rather than Popular-Cache, where $\beta$  can be solved by the bisection  method.
\item
Fixed-size interval access pattern: Popular-Cache is optimal under any  client request rate.
\end{itemize}
\subsubsection{$\lambda\to+\infty$}When $\lambda$ is relatively large,  the  cache allocation algorithms for different access patterns are  as follows.
\begin{itemize}
\item
Full access pattern: The average bandwidth consumption of  $V_i$ becomes $\frac{r}{f_B}\ln(\frac{L}{l_i})$, and  CCE-MP reduces to the proactive delivery mechanism which broadcasts the video periodically. We then have
\begin{align}
    \min_{\mathbf{l}}& ~~\sum\limits_{i=1}^{M}\frac{r}{f_B}\ln\left(\frac{L}{l_i}\right)\label{pro:FA_largerequest}\\
    \textup{s.t.}&  ~~(\ref{pro:cache_constraint1}), \left(\ref{pro:cache_constraint2}\right) \nonumber,
\end{align}
 which is also a convex problem. The optimal solution is to evenly allocate the cache capacity among all videos, i.e., $l_1=\ldots=l_M=\frac{C}{M}$. We term it as \textbf{Even-Cache}. The  total bandwidth consumption  in this  case is $\frac{Mr}{f_B}\ln(\frac{ML}{C})$, served as the ``upper bound" of the optimal scheme under any client request rate.
\item
Interval access pattern with random endpoints: The  bandwidth consumption of  $V_i$ is $\frac{r}{f_B}\ln(\frac{L}{l_i})$,  the same as that of the full access pattern. Therefore, Even-Cache is optimal.
\item
{\color{black} Fixed-size interval access pattern:  Based on {\bf Lemma \ref{lemma:fixedsizecache}}, the optimal cache allocation for this pattern is Popular-Cache under any client request rate. Therefore, different from the  access patterns with the same beginning where Even-Cache is optimal,  Popular-Cache is optimal for this pattern when request rate is relatively large. Note that when request rate becomes relatively large, the bandwidth consumptions of  $V_i$ under both the full access pattern and the interval access pattern with random endpoints become the same constant value  $\frac{r}{f_B}\ln(\frac{L}{l_i})$ eventually, yielding the Even-Cache allocation among all videos. However, this is not the case for the fixed-size interval access pattern. When $\lambda$ becomes relatively large, we have $E(T_i)=\sqrt{\frac{\pi{L}}{2r\lambda_i}}$ based on Eq. (\ref{equ:etix}), and the bandwidth consumption of   $V_i$ is $b_i^{FS}=\frac{L-l_i}{f_B}\sqrt{\frac{2r\lambda_i}{\pi{L}}}$,  still increasing with larger $\lambda_i$ value. Meanwhile, a larger popularity for a video implies a larger request rate. Therefore, a video with larger popularity should have larger cache storage size. Since it belongs to a fractional knapsack problem according to the proof part of Lemma \ref{lemma:fixedsizecache}}, Popular-Cache is optimal here.
\end{itemize}
\section{VoD Delivery in the Proactive  System}\label{sec_proactive}
Instead of minimizing the bandwidth consumption of VoD services  under the zero-delay  constraint  for  the proactive  system{\footnote{This problem reduces to Problem (\ref{pro:FA_largerequest}) with the minimum bandwidth consumption $\frac{Mr}{f_B}\ln(\frac{ML}{C})$.  In this case, the cache capacity is evenly allocated among all videos, and portion $dx$ at offset $x$ is broadcasted every $\frac{x}{r}$  time units, where $\frac{C}{M}\leq x\leq L$.}}, we jointly design the cache-bandwidth allocation and the multicast delivery   to minimize the average client waiting time under the total bandwidth constraint. {{\color{black}Note that for limited bandwidth-cache resource, the waiting time performance might not be guaranteed for each client. In this case, the typical  performance metric is to minimize the blocking probability if  client requests are blocked when their waiting times exceed their waiting tolerance, or to minimize  the average waiting time.  In this section, same as \cite{unicastcache,re07hybrid,pro14BackgroundPush}, we focus on the  average waiting time minimization problem{\footnote{As for the blocking probability minimization problem in the proactive system, the result is trivial, i.e., it is optimal to evenly allocate  cache and bandwidth among the most popular videos such that the waiting times  for these videos  just reach the tolerance.}}. Both full and interval access patterns are considered. }}

 Let $B$ denote the total bandwidth  and  $\mathbf{b}=[b_1,\ldots, b_M]$ denote the bandwidth allocation of each video, where $b_i$ is the allocated  bandwidth for broadcasting the $i$-th video periodically. The bandwidth constraint can be then written as $\sum_{i=1}^{M}{b_i}\leq B$.   Let $S_{pd}$ and $d_{i}(b_i,l_i, S_{pd})$ denote the adopted proactive delivery mechanism and the corresponding waiting time for the $i$-th video with allocated bandwidth $b_i$ and cache size $l_i$, respectively.  The average waiting time minimization problem can be then written in the following general  form:
\begin{align}\label{equ:general}
    \min_{\mathbf{b,l},S_{pd}}& ~~\sum\limits_{i=1}^{M}{p_i}d_{i}(b_i, l_i, S_{pd})\\
    \textup{s.t.}&  ~~\begin{cases}\sum\limits_{i=1}^{M}{b_i}\leq B, \sum\limits_{i=1}^{M}{l_i}\leq C,\\
    b_i\geq 0,   0\leq l_i\leq {L}, \forall i\in\{1,\ldots,M\}.\\
    \end{cases}\label{equ:constraint}
\end{align}

To minimize the average client waiting time, we first introduce the optimal  delivery mechanism given a certain cache-bandwidth allocation, and then develop the corresponding optimal cache-bandwidth allocation algorithm. To begin with, we  consider the traditional full access pattern.
 \subsection{Full Access Pattern}
{\color{black}For this pattern, the  greedy equal bandwidth broadcasting (GEBB) mechanism is optimal in the proactive system without client caching\cite{pro2001GEBB}.  In GEBB, the bandwidth for a certain video is equally divided into several subchannels and the video is also divided into different segments. Within each subchannel, a segment of the video is periodically broadcasted.  When requesting the video, the client starts buffering data on all subchannels. The division of the segments  meets the condition that a segment is entirely buffered (i.e., ready to display) right after the display of the previous segment is finished.} In  the following, we will  introduce a  client caching enabled GEBB (CCE-GEBB)  delivery mechanism, and  prove its optimality.
\begin{figure}[t]
\centering
\includegraphics[width=3.2in]{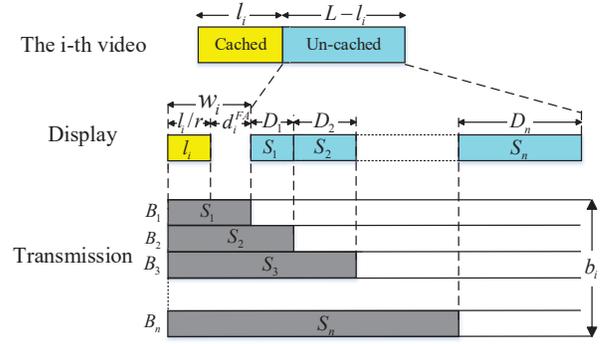}
\vspace{-1mm}
\caption{Illustration of the CCE-GEBB delivery mechanism for $V_i$ with allocated cache size $l_i$ and broadcast bandwidth $b_i$, where the grey blocks are the buffered data after the client request time.}\label{fig:GEBB}
\vspace{-3mm}
\end{figure}
\subsubsection{\textbf{ CCE-GEBB Delivery Mechanism}} CCE-GEBB  divides each video into cached and uncached parts, where the cached part is pre-cached at the client side  and the uncached part is  periodically broadcasted via a given  bandwidth. Taking $V_i$ with bandwidth $b_i$ and cache size $l_i$ for example, the uncached part with length $L-l_i$ is divided into $n$ segments as illustrated in Fig. \ref{fig:GEBB}. The $k$-th segment is of video length $S_k$ and duration $D_k$, which is repeatedly broadcasted over a channel of bandwidth $B_k=b_i/n$. The property of the optimally-structured broadcasting protocol is that  the data of the $k$-th segment has been entirely buffered right after the end of the previous segment is displayed, i.e.,
\begin{equation}
(w_i+\sum\limits_{j=1}^{k-1}D_j){B_k}{f_B}=rD_k,\quad\text{for }~ k=1,2,\ldots,n ,
\end{equation}
where $w_i$ is the waiting time to successfully display the uncached  part of $V_i$ after the  client request time. We have
\begin{equation}
\prod_{k=1}^{n}\left(\frac{B_kf_B}{r}\!+\!1\right)\!=\!\frac{w_i\!+\!\sum\limits_{j=1}^{n}D_j}{w_i}\!=\!\frac{w_i+(L-l_i)/r}{w_i},
\end{equation}
and
\begin{equation}
w_i=\frac{L-l_i}{r}\left[\left(1+\frac{Bf_B}{nr}\right)^n-1\right].
\end{equation}
The minimum $w_i$ can be achieved when $n\rightarrow +\infty$, i.e,
\begin{equation}
w_{i}=\frac{L-l_i}{r(e^{\frac{f_B}{r}b_i}-1)}.
\end{equation}
Since the pre-cached part of the video can be displayed locally, the waiting time that clients experience for $V_i$  is
\begin{equation}
d_{i}^{FA}=\left(w_{i}-\frac{l_i}{r}\right)^+=\left(\frac{L-e^{\frac{f_B}{r}b_i}l_i}{r(e^{\frac{f_B}{r}b_i}-1)}\right)^+.
\end{equation}
When $e^{\frac{f_B}{r}b_i}l_i=L$, the first segment $S_1$ has  been buffered right after the end of local cache is displayed (i.e., $w_i={l_i}/{r}$), yielding zero-delay for the $i$-th video. Note that there is no need to waste extra cache-bandwidth resource achieving $e^{\frac{f_B}{r}b_i}l_i>L$ while maintaining the same viewing experience,  then we have
\begin{equation}
d_{i}^{FA}=\frac{L-e^{\frac{f_B}{r}b_i}l_i}{r(e^{\frac{f_B}{r}b_i}-1)} \quad \textup{s.t.} ~ e^{\frac{f_B}{r}b_i}l_i\leq L.
\end{equation}
\begin{proposition}\label{pro:CCE-GEBB}
CCE-GEBB is the optimal proactive delivery mechanism to minimize the waiting time for the $i$-th video with  bandwidth $b_i$ and cache size $l_i$.
\end{proposition}
\begin{IEEEproof}
See Appendix D.
\end{IEEEproof}
\subsubsection{\textbf{Cache-bandwidth Allocation}} Given the developed CCE-GEBB delivery mechanism, the  cache-bandwidth allocation  problem becomes
\begin{align}\label{equ:generalproblem}
    \min_{\mathbf{b,l}}& ~~\sum\limits_{i=1}^{M}{p_i}\frac{L-e^{\frac{f_B}{r}b_i}l_i}{r(e^{\frac{f_B}{r}b_i}-1)}\\
    \textup{s.t.}&  ~~\begin{cases}\small{\sum\limits_{i=1}^{M}{b_i}\leq B,  \sum\limits_{i=1}^{M}{l_i}\leq C},\\
   \small{b_i\!\geq 0, 0\leq l_i\leq Le^{-\frac{f_B}{r}b_i}, \forall i\in\{1,\ldots,M\}.}
    \end{cases}
\end{align}

{\color{black}We first introduce the following proposition about the structure of the optimal solution.}
\begin{proposition}\label{pro:fa_structure}
If $V_k$ has  non-zero cache  allocated  and experiences non-zero delay, i.e., $0<l_k<Le^{-\frac{f_B}{r}b_k}$, then
 \begin{enumerate}[a)]
\item $V_1$ to $V_k$ with larger popularity experience zero-delay while  the remaining $M-k$ videos with smaller popularity have non-zero delay and no cache storage allocated.
\item   For the first $k-1$ videos with zero-delay, the storage and the bandwidth are evenly allocated, i.e., $b_1=\ldots=b_{k-1}$ and $l_1=\ldots=l_{k-1}$.
\end{enumerate}
\end{proposition}
\begin{IEEEproof}
See Appendix  E.
\end{IEEEproof}

Based on the structure of the optimal allocation in \textbf{Proposition \ref{pro:fa_structure}},  we then have the following lemma to solve the cache-bandwidth allocation problem.
\begin{lemma}\label{lemma:fa_optimal}
For the full access pattern, the  cache-bandwidth allocation problem of $2M$ variables can be simplified to a one-dimensional search of the first cache size $l_1$
\begin{equation}
(\mathbf{l^*,b^*})=\arg\max_{l_1}\phi(l_1),
\end{equation}
where  $l_1\in[L/M, \min(L,C)]$ and $\phi(l_1)$ is the average waiting time  in terms of $l_1$ when the cache allocation is
 \begin{equation}
 l_i=\begin{cases}l_1 \quad &\text{if } 1 \leq i \leq k-1,\\
 C\bmod l_1 \quad &\text{if }  i= k,\\
 0 \quad &\text{if } k+1 \leq i \leq M,
 \end{cases}
\end{equation}
in which $k=\lfloor\frac{C}{l_1}\rfloor+1$ is the threshold video number which has  cache storage allocated.

If $l_1<C$, the  bandwidth allocation becomes
\begin{align}
 b_i=\begin{cases}&\frac{r}{f_B}\ln{\frac{L}{l_1}}  \quad~~~~~~ ~~~~~~~~~~~\text{if } 1 \leq i \leq k-1,\\
 &\!\frac{r}{f_B}\ln\!\big(\frac{2\!+\!{p_i\beta(L-l_{i})}\!+\!\sqrt{(p_i\beta(L-l_{i}))^2\!+\!4{p_i\beta(L-l_{i})}}}{2}\big) \\
  &~~~~~~~~~~~~~~~~~~~~~~~~~~~~\quad \text{if } k \leq i \leq M,
 \end{cases}
 \end{align}
 where $\beta$  meets  the {bandwidth} constraint $\sum\limits_{i=1}^{M}b_i=B$.

 If $l_1=C\leq L$,   the cache capacity is allocated to the first video and we have, for $i=1,2,\ldots, M$,
 \begin{equation}
 \small{\!b_i\!=\!\frac{r}{f_B}\!\ln\!\Big(\frac{2\!+\!{p_i\beta(L-l_{i})}\!+\!\sqrt{(p_i\beta(L-l_{i}))^2\!+\!4{p_i\beta(L-l_{i})}}}{2}\!\Big)}
 \end{equation}
 \end{lemma}
\begin{IEEEproof}
See Appendix F.
\end{IEEEproof}

 If and only if the total bandwidth meets $B\geq \frac{Mr}{f_B}\ln(\frac{ML}{C})$, zero-delay  can be achieved for all videos, which is consistent with the extreme case analysis in Sec. \ref{subsec:extreme}.
\subsection{Interval Access Pattern with Random Endpoints}
For the interval access pattern with uniformly distributed endpoints, we first introduce the proactive delivery mechanism and then derive the optimal cache-bandwidth allocation.
\subsubsection{\textbf{CCE-GEBB Delivery Mechanism}}  Clients who finish the watching of the $i$-th video before position $l_i$ experience zero-delay, and the corresponding probability is ${l_i}/{L}$ due to the uniform distribution of endpoints.   For the remaining  clients interested in the uncached part, the waiting time is  $d_{i}^{FA}$ as indicated in Fig. \ref{fig:GEBB}. Since CCE-GEBB minimizes $d_{i}^{FA}$, it also achieves the minimum waiting time under  this access pattern.  Based on CCE-GEBB,  the average waiting time of  $V_i$ is
\begin{align}
d_{i}^{RE}=\frac{L-l_i}{L}d_{i}^{FA}=\frac{L-l_i}{L}\left(\frac{L-l_i}{r(e^{\frac{f_B}{r}b_i}-1)}-\frac{l_i}{r}\right),
\end{align}
\subsubsection{\textbf{Cache-bandwidth Allocation}}
Let $x_i={1}/{e^{\frac{f_B}{r}b_i}}$ for $i\in\{1,\ldots,M\}$,    and note that the cache-bandwidth allocation problem can be rewritten as
\begin{align}
    \min_{\mathbf{x,l}}& ~~\sum\limits_{i=1}^{M}{p_i}\frac{L-l_i}{rL}\left(-\frac{L-l_i}{x_i-1}-L\right)\label{pro_rn}\\
    \textup{s.t.}&  ~~\begin{cases}\sum\limits_{i=1}^{M}\ln{\frac{1}{x_i}}\leq \frac{Bf_B}{r},
    \sum\limits_{i=1}^{M}{l_i}\leq C,\\
    0 < x_i\leq 1, 0\leq l_i\leq {L}x_i, \forall i\in\{1,\ldots,M\}.\\
    \end{cases}\nonumber
\end{align}
\begin{proposition}
Problem (\ref{pro_rn}) is convex.
\end{proposition}
\begin{IEEEproof}
The hessian matrix of $d_{i}^{RE}$ becomes
\begin{equation}\nonumber
\small{
\mathbf{H}\!=\!\begin{bmatrix}\frac{\partial^2d_{i}^{RE}}{\partial{x_i}^2}&\frac{\partial^2d_{i}^{RE}}{\partial x_i\partial l_i}\\\frac{\partial^2d_{i}^{RE}}{\partial l_i\partial x_i}&\frac{\partial^2d_{i}^{RE}}{\partial {l_i}^2} \end{bmatrix}
\!=\!\frac{2}{r{\color{black}L}(1-x_i)}\begin{bmatrix}\frac{L-l_i}{1-x_i}\\{-1}\end{bmatrix}\begin{bmatrix}\frac{L-l_i}{1-x_i}&{-1}\end{bmatrix}^T\!\geq\! 0,}
\end{equation}
thus $d_{i}^{RE}$ is convex  in $x_i$ and $l_i$, and the objective function is convex along with convex constraints. Hence, the problem is convex and can be  solved by the interior-point method.
\end{IEEEproof}

Similar to \textbf{Proposition \ref{pro:fa_structure}}, we also have the following statement for this access pattern.
\begin{proposition}
If $V_k$ experiences zero-delay, then
 \begin{enumerate}[a)]
 \item $V_1$ to $V_{k-1}$ experience zero-delay.
  \item For the videos with zero-delay, the cache size and the bandwidth are  evenly allocated.
 \end{enumerate}
\end{proposition}
\subsection{Access Pattern with Downloading Demand}
\begin{table}
\caption{Default parameter settings}
\vspace{-3mm}
\label{table:setting}
\begin{center}
\begin{tabular}{|c|l|c|}
\hline
\textbf{Parameter} & \textbf{Description}&\textbf{Value} \\ \hline
$M$ & Number of videos & $200$\\ \hline
$\alpha$ & Zipf parameter & $0.8$\\ \hline
$L$ &Video length & $150$ MB ($10$ minutes) \\ \hline
$r$ &Video bitrate & $2$ Mbps \\ \hline
$f_B$&Bandwidth efficiency &$4$ bps/Hz \\ \hline
$C$&Cache size& $3000$ MB\\ \hline
$\lambda$ &Client request rate& ${\color{black}0.5}$ $s^{-1}$\\ \hline
\end{tabular}
\end{center}
\end{table}
\begin{table*}[ht]
\renewcommand{\arraystretch}{1.1}
\caption{Illustration of different schemes}
\label{table:schemes}
\begin{center}
\begin{tabular}{|l|l|c|c|c|}
\hline
\textbf{Scheme}&\textbf{System}&\textbf{Delivery Mechanism}&\textbf{Cache Allocation}&\textbf{Bandwidth Allocation} \\ \hline
R-optimal&\multirow{5}{*}{reactive}&CCE-MP& optimal& ---\\   \cline{1-1}\cline{3-5}
R-popularCache& &CCE-MP&Popular-Cache&---\\  \cline{1-1}\cline{3-5}
R-evenCache& &CCE-MP& Even-Cache&---\\  \cline{1-1}\cline{3-5}
Batch\cite{Niulab}& &Batch&\cite{Niulab}&---\\  \cline{1-1}\cline{3-5}
PAB-MP\cite{Melbourne}& &PAB-MP&\cite{Melbourne}&---\\ \hline
P-optimal&\multirow{5}{*}{proactive}& CCE-GEBB& optimal& optimal\\  \cline{1-1}\cline{3-5}
P-popularCache&&CCE-GEBB&Popular-Cache&optimal\\  \cline{1-1}\cline{3-5}
P-evenCache&&CCE-GEBB& Even-Cache&optimal\\  \cline{1-1}\cline{3-5}
P-even&&CCE-GEBB&Even-Cache&evenly allocated\\  \cline{1-1}\cline{3-5}
P-noStorage&&GEBB&---&optimal\\ \hline
\end{tabular}
\end{center}
\end{table*}

In this subsection, we consider the downloading-demand access pattern where each client selectively watches interested video clips  until the desired video is fully saved{\footnote{The downloading-demand access pattern is not studied for the reactive system in Sec. \ref{sec:reactive} since the considered zero-delay constraint is not practical for this  pattern. Instead, this pattern can be investigated for the reactive system given a maximum downloading time constraint.}}.   In this case, the client waiting time  reduces to the  downloading time.

\subsubsection{\textbf{CCE-GEBB Delivery Mechanism}}  Under this  access pattern, no part in a video has a higher timing priority than others in the transmission and each bit of the video should be sent at the same frequency.  In this case,  the total number of subchannels in  CCE-GEBB becomes 1 (i.e., $n=1$ in Fig. \ref{fig:GEBB}), and CCE-GEBB reduces to the traditional broadcast carousel where the uncached data of $V_i$ is  cyclically transmitted via one subchannel with bandwidth $b_i$.   The downloading time of  $V_i$ with bandwidth $b_i$ and cache size $l_i$ is
\begin{equation}
{d}_{i}^{DD}=\frac{L-l_i}{f_Bb_i}.
\end{equation}
\subsubsection{\textbf{Cache-bandwidth Allocation}}The resource allocation problem  becomes
\begin{equation}\label{pro:dd}
\begin{split}
    \min_{\mathbf{b,l}}& ~~\sum\limits_{i=1}^{M}p_i\frac{L-l_i}{f_Bb_i}\\
    \textup{s.t.}&  ~~(\ref{equ:constraint}).
\end{split}
\end{equation}
And we have the following {\color{black}proposition}.
\begin{proposition}\label{theorem:FD_cache}
 The optimal cache allocation for  this access pattern is Popular-Cache, i.e.,
 \begin{equation}
 l_i=\begin{cases}L \quad &\text{if } i\leq k,\\
C\bmod L \quad &\text{if }  i=k,\\
 0 \quad &\text{if }  i>k,
 \end{cases} \quad
\end{equation}
 where $k=\lfloor{C}/{L}\rfloor+1$. The corresponding  bandwidth allocation is
\begin{equation}
b_i=\frac{B\sqrt{p_i(L-{l_i})}}{\sum\limits_{j=1}^{M}\sqrt{p_j(L-{l_j})}},\quad \text{for } i=1,2,\ldots,M.
\end{equation}
\end{proposition}
\begin{IEEEproof}
See Appendix G.
\end{IEEEproof}

\section{Performance Evaluation}\label{sec:numerical}
{\color{black}In this section, simulations are provided to validate the performance gain of the proposed schemes in  both reactive and proactive systems.  The default parameter settings are shown in  Table \ref{table:setting}. In  our simulation, the number of videos in the library is taken as 200.  Each video is of  bitrate $2$ Mbps and duration 10 minutes \cite{unicastcache}. The popularity of each video is distributed according to a Zipf law of parameter $\alpha$\cite{Zipf}, where $\alpha$ governs the skewness of the popularity. The popularity is uniform over videos for $\alpha=0$, and becomes more skewed as $\alpha$ grows. We select  $\alpha=0.8$ as the default value\cite{Melbourne}, where $47\%$ client requests concentrate on the $10\%$  popular videos.  The  client cache size is  $3000$ MB ($2.93$ GB), which  is reasonable for smart devices with increasing cache storage size (e.g., 16 GB).

Table \ref{table:schemes} illustrates the evaluated schemes adopted in the simulation. In Batch \cite{Niulab}, multiple  client requests for the same video that arrive within a batching window (i.e., the local displaying period due to prefix cache) are grouped and served via a single multicast transmission.  In the PAB-MP scheme \cite{Melbourne}, in addition to the prepopulation assisted batching, clients can join an ongoing multicast stream and multicast patching streams are scheduled to patch the missing parts.}
\subsection{Reactive System}
\begin{figure}[t]
\centering
 \includegraphics[height=2.6in]{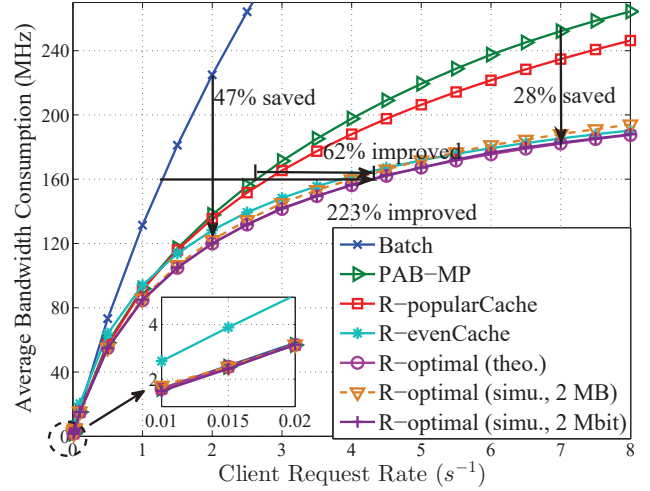}
 \caption{Impact of  request rate $\lambda$ on the average  bandwidth consumptions  under the full access pattern.}
 \label{fig:cp_lambda}
  \vspace{-3mm}
\end{figure}

\begin{figure}[t]
\centering
 \includegraphics[height=2.6in]{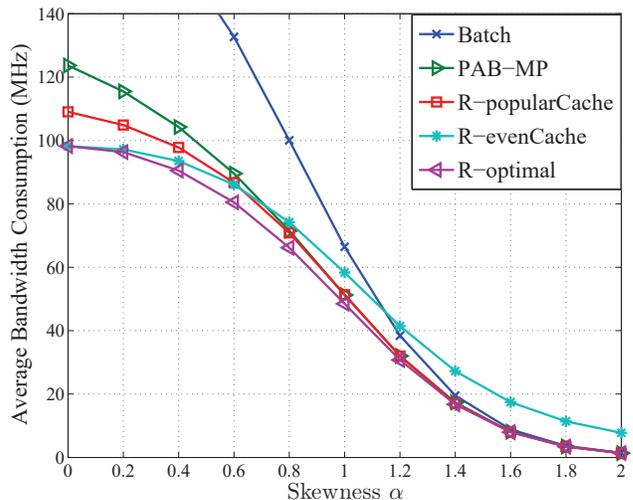}
 \caption{Impact of Zipf parameter $\alpha$ on the  average bandwidth consumptions under the full access pattern.}
 \label{fig:cp_zipf}
  \vspace{-3mm}
\end{figure}

\begin{figure}[t]
\centering
 \includegraphics[height=2.6in]{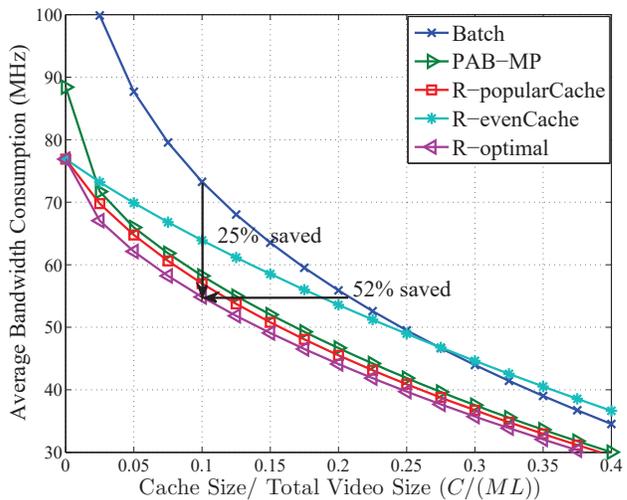}
 \caption{Impact of cache size on the  average bandwidth consumption under the full access pattern.}
 \label{fig:cp_size}
  \vspace{-3mm}
\end{figure}

\begin{figure}[t]
\centering
 \includegraphics[height=2.6in]{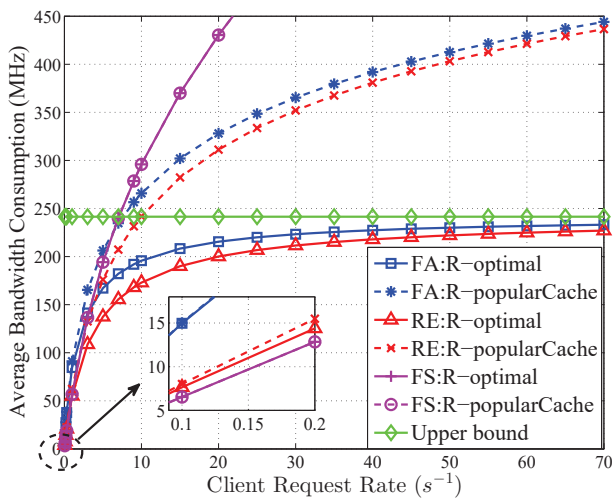}
 \caption{Impact of $\lambda$ on the average  bandwidth consumptions of different access patterns with  $D=4$ min.}
 \label{fig:cp_3_lambda}
  \vspace{-3mm}
\end{figure}

 For the reactive system, the impacts of the client request rate,  the Zipf parameter, the cache size {\color{black}and the number of videos} on the average bandwidth consumptions of different schemes are illustrated in Figs. \ref{fig:cp_lambda}, \ref{fig:cp_zipf}, \ref{fig:cp_size} and \ref{fig:cp_number}, respectively. In addition,  the impact of the access pattern is  shown in Fig. \ref{fig:cp_3_lambda}.

\textbf{Impact of client request rate}: The impact of client request rate $\lambda$ on the {\color{black}average}  bandwidth consumption under the  full access pattern is presented in Fig. \ref{fig:cp_lambda}, {\color{black}where R-optimal (simu., 2 MB) and R-optimal (simu., 2 Mbit)  stand for the practical case that video chunks are of 2 MB ($75$ chunks) and 2 Mbit ($600$ chunks), respectively, rather than the arbitrary small size in R-optimal (theo.). The smaller the video chunk size, the smaller the performance degragation compared to R-optimal (theo.).  Note that the simulation result of  R-optimal (simu., 2 Mbit)  achieves  nearly the same performance with R-optimal (theo.),  hence the chunk video size  $2$  Mbit is  adopted in the following simulations for the reactive system.}
 In addition, the performances of Batch, PAB-MP, R-popularCache  and the proposed R-optimal  are nearly the same under relatively low request rates (e.g., $\lambda=0.01$).  Since there is little chance to merge multiple client requests at that low request rate,  the server  responds to almost all client requests via unicast. In this case, it is optimal to simply cache the most popular videos. As $\lambda$ increases, by buffering one ongoing stream and later receiving a corresponding multicast patching stream, PAB-MP outperforms Batch where clients join no ongoing streams. However, both Batch and PAB-MP suffer  significant  performance losses compared to R-optimal since R-optimal utilizes every desired part of ongoing streams.  For instance,  up to $47\%$ (or $28\%$)  bandwidth saving can be achieved by R-optimal  compared to Batch (or PAB-MP) at $\lambda=2$ (or $7$), {\color{black} and $223\%$ (or $62\%$) more requests can be supported by R-optimal compared to Batch (or PAB-MP) at bandwidth consumption $160$ MHz}. Moreover, R-evenCache  has almost the same performance with R-optimal  when $\lambda\geq7$, which coincides with the extreme case analysis in Sec. \ref{subsec:extreme}, i.e.,  it is optimal to evenly allocate the cache capacity among all  videos under relatively high request rates.

\textbf{Impact of Zipf parameter}: Fig. \ref{fig:cp_zipf} illustrates the  average  bandwidth consumptions of various schemes vs. $\alpha$ under the full access pattern. When $\alpha=0$,  the popularity is uniformly distributed and  R-optimal reduces to R-evenCache. As $\alpha$ increases, more  requests  concentrate on the  first few videos,  resulting in less bandwidth consumption for all schemes. Note that R-evenCache, which  employs CCE-MP,  is even worse than Batch for $\alpha>1.2$ since the adopted  Even-Cache   ignores the popularity property.  In addition,  R-popularCache  performs nearly the same as R-optimal  for that the first few cached videos  dominate most requests for large $\alpha$.

\textbf{Impact of cache size}: As shown in Fig. \ref{fig:cp_size}, the {\color{black}average}  bandwidth consumptions of all schemes decrease with increasing cache size since a larger cache  size provides larger local-cache and multicast  gains.  Note that R-popularCache outperforms R-evenCache under the settings $\lambda=0.5$  (low request rate) and $\alpha=0.8$ (highly skewed popularity). {\color{black}Compared with Batch, R-optimal saves $25\%$ bandwidth consumption at the same cache size $0.1ML$, and reduces $52\%$ cache consumption while achieving the same bandwidth consumption.} {\color{black}As the cache size increases, the performance gap between R-optimal and Batch becomes smaller, which coincides with \textbf{Lemma 1}.} {\color{black} Moreover, R-optimal in Fig. \ref{fig:cp_size} indicates the minimum cache-bandwidth resource required for supporting a certain client request rate, e.g., ($0.1ML$, $55$ MHz) and ($0.2ML$, $44$ MHz), which can be used as a guideline for VoD services with client caching.}

{\color{black}
\begin{figure}[t]
\centering
 \includegraphics[height=2.6in]{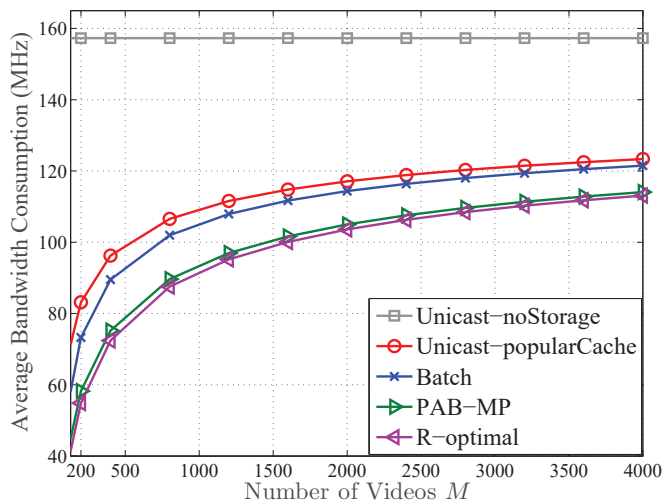}
 \caption{Impact of the number of videos on the average bandwidth consumption under the full access pattern.}
 \label{fig:cp_number}
  \vspace{-3mm}
\end{figure}
}

{\color{black}\textbf{Impact of the total number of videos:} The impact of the  number of videos on the bandwidth consumptions of different schemes is illustrated in Fig.  \ref{fig:cp_number}, where the Unicast-popularCache scheme caches only the most popular videos  and serves the remaining video requests via unicasting.  With the increase of the number of videos,  the gap between Unicast-popularCache and Batch become relatively small.  The reason is that the cache size and the video popularity for each video becomes smaller with larger $M$, and fewer client requests are batched for the videos with smaller cache size and smaller video popularity through a single transmission. In this case,  the performance of Batch would reduce to that of  Unicast-popularCache eventually.    However,  PAB-MP and R-optimal  still have notable bandwidth saving compared to Batch even with $M=4000$, since both schemes utilizes the patching method to  exploit the ongoing streams. When $M=4000$, the cache storage can only cache $0.5\%$ of the total videos. However, Unicast-popularCache still saves $21.6\%$ bandwidth consumption compared to Unicast-noStorage. The reason is that  the most popular $0.5$ percent of total videos accounts for $21.6\%$ of the total requests when $M=4000$, showing the effectiveness of client caching.}

\textbf{Impact of different access patterns}: Fig. \ref{fig:cp_3_lambda} illustrates the  average bandwidth consumptions of different access patterns vs. the client request rate, where ``FA", ``RE" and ``FS" represent the full access pattern, the interval access pattern with random endpoints and the fixed-size interval access pattern, respectively.  Under  relatively low client request rates (e.g., $\lambda=0.1$), client requests could not be merged and the server responds  by unicast. Therefore,  R-popularCache is optimal under ``FA"  while it suffers a performance loss under ``RE".  As $\lambda$ increases, the average bandwidth consumption of ``FS" is even larger than that of ``FA" since more multicast opportunities could be exploited by access patterns with the same beginning. Meanwhile, {\color{black}the performance gap of R-optimal between ``FA" and ``RE" becomes smaller},  and both patterns approach the ``upper bound" with increasing $\lambda$.  Furthermore, R-popularCache is optimal  under ``FS" for all request rates. Therefore, the numerical results are consistent with  the  theoretical analyses in Sec. \ref{sec:reactive}.

\subsection{Proactive system}
For the proactive system, we aim to minimize the average client waiting time under the total bandwidth constraint. The impacts of the cache size, the Zipf parameter and the access pattern are  described as follows, where the evaluated total bandwidth  is $130$ MHz.
\begin{figure}[t]
\centering
\includegraphics[height=2.6in]{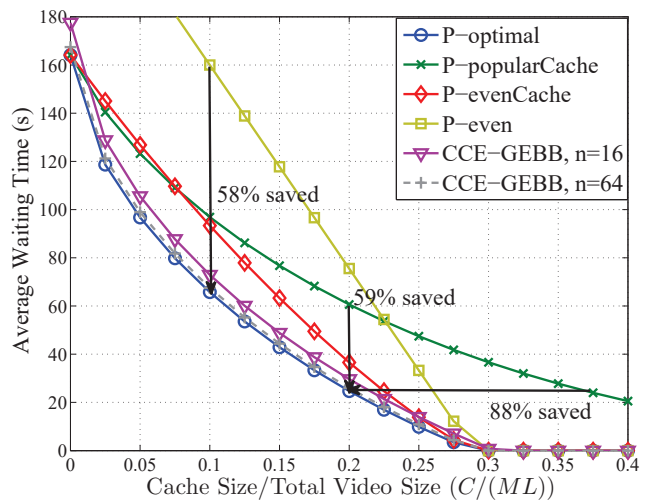}
 \caption{Impact of cache size on the average waiting times of different schemes under the full access pattern.}
 \label{fig:op_fa_size}
  \vspace{-3mm}
\end{figure}
\begin{figure}[t]
\centering
\includegraphics[height=2.6in]{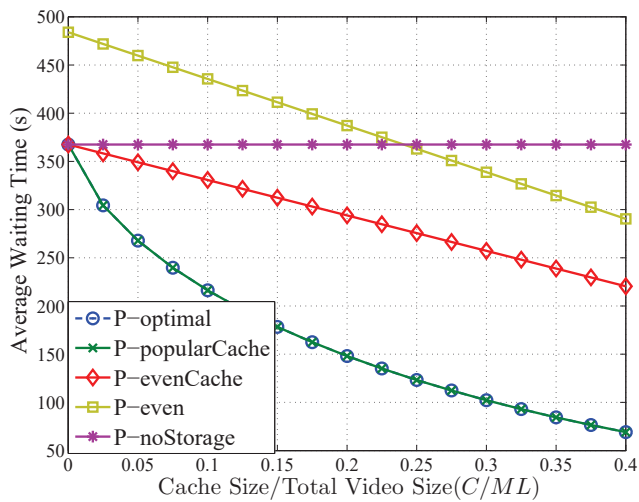}
 \caption{Average waiting times of different schemes vs cache size under the downloading-demand access pattern.}
 \label{fig:pro_download_storage}
  \vspace{-3mm}
\end{figure}

\textbf{Impact of cache size}: The impact of the cache size on the average waiting times of different schemes under the full access pattern is illustrated in Fig. \ref{fig:op_fa_size}, where P-popularCache caches only the most popular videos  while P-evenCache and P-even evenly cache the prefixes of all videos.  As indicated in Table \ref{table:schemes}, the bandwidth in P-evenCache is optimally allocated given the Even-Cache allocation while it is still evenly allocated in P-even.  ``CCE-GEBB, n=16" (or n=64) stands for the practical scenario where  each video is transmitted over 16 (or 64) subchannels rather than infinite subchannels. Note that the more subchannels  are allocated for each video, the less performance degradation is obtained compared to the infinite case, e.g.,  ``CCE-GEBB, n=64" achieves nearly optimal. In addition,  P-evenCache and P-even perform worse than P-popularCache under small cache size settings (e.g., $C<0.075ML$) while P-evenCache achieves nearly optimal under large cache size settings (e.g., $C\geq0.25ML$). This is due to the fact that  the bandwidth and the cache capacity are evenly allocated among videos with zero-delay, and a larger cache size yields more videos with zero-delay. Compared with R-popularCache (or R-evenCache), the proposed R-optimal reduces $59\%$ (or $58\%$)  average waiting time at the same cache size $0.2ML$ (or $0.1ML$).  {\color{black}Moreover, Fig. \ref{fig:op_fa_size}  provides useful insights in choosing the appropriate cache size to meet the average waiting time constraint, e.g., R-optimal with cache size $0.2ML$ saves $88\%$ cache consumption compared to P-popularCache while meeting  the same average waiting time constraint (i.e., $25$ s)}. Similar results can also be expected  for the interval access pattern with random endpoints.

\begin{figure}[t]
\centering
\includegraphics[height=2.6in]{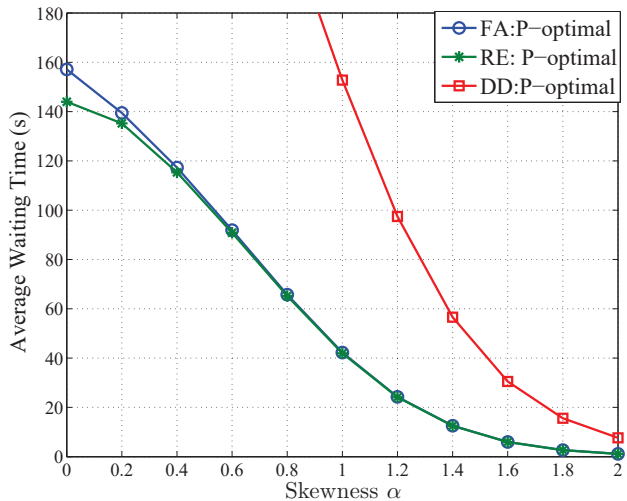}
 \caption{Impact of Zipf parameter on the average waiting times of three different access patterns.}
 \label{fig:pro_3_zipf}
  \vspace{-3mm}
\end{figure}

Fig. \ref{fig:pro_download_storage} illustrates the impact of the cache size on the average waiting time under the access pattern with downloading demand. Unlike the access patterns starting from the same beginning, the optimal proactive delivery mechanism under this pattern is CCE-GEBB with $n=1$ (i.e., the traditional broadcast carousel), and the   cache allocation  is to cache the most popular videos only. {\color{black}When $C/(ML)=0.4$, the average waiting time of the optimal scheme is still $67$ seconds, while it is  zero  under the full access pattern in Fig. \ref{fig:op_fa_size}}.

\textbf{Impact of Zipf parameter}:  Fig. \ref{fig:pro_3_zipf} illustrates the average waiting times of different access  patterns v.s. the Zipf parameter $\alpha$, where ``FA", ``RE" and ``DD" represent the full access pattern, the interval access pattern with random endpoints and the access pattern with downloading demand, respectively.   As  $\alpha$ increases, the  performance gap between  ``FA" and ``RE"  becomes negligible since the optimal schemes under both access patterns aim to provide zero-delay for videos with larger popularity.  Meanwhile, {\color{black}The performance gap between ``DD" and other two access patterns becomes much smaller with increasing $\alpha$}. This is due to the fact that  Popular-Cache is optimal for ``DD",  and  most client requests concentrate on the popular contents already cached  at client side for large $\alpha$.

\begin{figure}[t]
\centering
\includegraphics[height=2.6in]{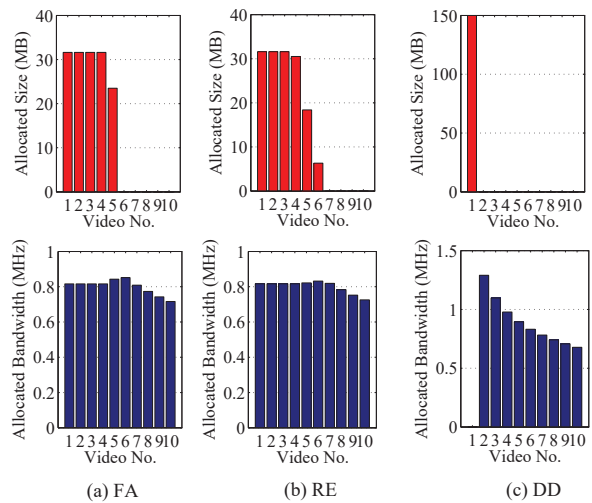}
\caption{Cache and bandwidth allocations under three different access patterns with $M=10$ and $B=8$ MHz.}
\label{fig:allocation}
 \vspace{-3mm}
\end{figure}

\textbf{Impact of different access patterns}: The cache-bandwidth allocations under  ``FA", ``RE" and ``DD"  are illustrated in Figs. \ref{fig:allocation}(a), \ref{fig:allocation}(b) and \ref{fig:allocation}(c),  respectively. For  popular videos with zero-delay under ``FA" and ``RE" {\color{black}(i.e., $V_1$ to $V_4$ in Fig. \ref{fig:allocation}(a) and  $V_1$ to $V_3$ in Fig. \ref{fig:allocation}(b))}, the cache and the bandwidth are evenly allocated. Meanwhile, the videos with zero-delay are entirely cached under ``DD"  {\color{black}(i.e., $V_1$ in Fig. \ref{fig:allocation}(c))}.
For ``FA",  at most one video {\color{black}( i.e., $V_5$ in Fig. \ref{fig:allocation}(a))} with non-zero delay has  cache  allocated due to the greedy property of the solution for  fractional knapsack problems, while several videos {\color{black}(i.e., $V_4$ to $V_6$ in Fig. \ref{fig:allocation}(b))} with non-zero delay could have cache  allocated under ``RE". In addition, Popular-Cache is optimal under ``DD" in Fig. \ref{fig:allocation}(c). Therefore, the numerical results are consistent with  the theoretical analyses in Sec. \ref{sec_proactive}.

\section{Conclusion}\label{sec:conclusion}
This paper has investigated the optimal joint resource allocation and multicast delivery schemes for VoD services  in reactive and proactive systems with client caching. Both  full and interval access patterns have been considered.   For the reactive  system,  we have developed a joint cache allocation and multicast delivery scheme to minimize the average bandwidth consumption under the zero-delay constraint. We observe that in addition to the video popularity, the cache allocation algorithm also relies on the client request rate and the access pattern.  For the proactive system,  we have jointly designed the cache-bandwidth allocation algorithm and the CCE-GEBB delivery mechanism to minimize the average waiting time under the total bandwidth constraint. Note that CCE-GEBB with infinite subchannels  is optimal  for both the full access pattern and the interval access pattern with random endpoints,  and the cache capacity is evenly allocated among videos with zero-delay. Meanwhile,  CCE-GEBB  with only one subchannel is optimal for the access pattern with downloading demand, in which case the optimal cache allocation  is to cache the most popular contents entirely. {\color{black}These results can be used as a guideline for the VoD network with client caching, e.g., the required  minimum bandwidth-cache resource  under  a certain client request rate in the reactive system, or under a certain average waiting time constraint in the proactive system.}
\appendices
\section*{Appendix A: Proof of Proposition~\ref{pro:CCE-MP}}\label{app:CCE-MP}
 Assuming by contradiction that a part of a desired video is multicasted  before the time of display, then the average bandwidth consumption could be increased since the following requests for the same  part before that time are not benefited by the transmission and more data needs to be sent.  Also if a client does not start buffering the desired uncached data from an ongoing stream right after the request time, then the useful data in the ongoing stream might not be fully utilized, resulting in  extra data transmission for that client. Therefore,  CCE-MP is  optimal  to minimize the  average bandwidth consumption given a certain cache allocation under any access pattern.
 {\color{black}\section*{Appendix B: Proof of Lemma~\ref{lemma:gap}}
 The bandwidth consumption of $V_i$ under CCE-MP can be rewritten as
\begin{equation}
b_{i}^{FA}=\frac{r}{f_B}\ln\left(\frac{L-l_i}{l_i+\frac{r}{\lambda_i}}+1\right)=\frac{r}{f_B}\ln\left(\frac{f_Bb_{i,{batch}}}{r}+1\right).
\end{equation}
We then have
\begin{equation}
\frac{\partial \frac{b_{i,batch}-b_{i}^{FA}}{b_{i,batch}}}{\partial {b_{i,{batch}}}}=\frac{r}{f_B}\frac{\ln\left(\frac{f_Bb_{i,{batch}}}{r}+1\right)-\frac{\frac{f_Bb_{i,batch}}{r}}{{\frac{f_Bb_{i,batch}}{r}}+1}}{(b_{i,batch})^2}.
\end{equation}
Denote $g(t)=\ln(t+1)-\frac{t}{t+1}$ where $t\geq 0$, we then have $g'(t)=\frac{t}{(t+1)^2}\geq 0$. Therefore, $g(t)\geq g(0)=0$ and we have
\begin{equation}
\frac{\partial \frac{b_{i,batch}-b_{i}^{FA}}{b_{i,batch}}}{\partial {b_{i,{batch}}}}=\frac{r}{f_B}\frac{g(\frac{f_Bb_{i,batch}}{r})}{(b_{i,batch})^2}\geq 0.
\end{equation}
Therefore, $\frac{b_{i,batch}-b_{i}^{FA}}{b_{i,batch}}$ increasing with larger  $b_{i,batch}$ values. Since $b_{i,batch}$ decreases with larger cache size $l_i$, then $\frac{b_{i,batch}-b_{i}^{FA}}{b_{i,batch}}$ also decreases with larger cache size $l_i$.}
\section*{Appendix C: Proof of Lemma~\ref{lemma:waterfill}}\label{app:waterfill}
The second derivative of $b_{i}^{FA}$ is
\begin{equation}
\frac{\partial^2{b_{i}^{FA}}}{\partial{{l_i}^2}}=\frac{r}{f_B(l_i+\frac{r}{\lambda_i})^2}> 0,
\end{equation}
 thus  $b_i^{FA}$ is convex in $l_i$ and  $\sum_{i=1}^M{b_i^{FA}}$ is also convex. Therefore, (\ref{pro:cache_fa}) is  a convex problem.  Consider the  Lagrangian
\begin{equation}
{\cal{L}^{FA}}=\sum\limits_{i=1}^{M}b_{i}^{FA}+\mu(\sum\limits_{i=1}^{M}l_i-C),
\end{equation}
where $\mu$ is the Lagrange multiplier.  The Karush-Kuhn-Tucker (KKT) condition for the optimality of a cache allocation is
\begin{equation}
\frac{\partial\cal{L^{FA}}}{\partial{l_i}}=-\frac{r}{f_B(l_i+\frac{r}{\lambda_i})}+\mu\begin{cases}=0 \quad \text{if } 0<l_i<L,\\
\geq 0 \quad \text{if } l_i=0,\\
\leq 0 \quad \text{if } l_i=L.
\end{cases}
\end{equation}
Let {\color{black}$\beta=1/({f_B\mu})$} and $x^+=\max{(x,0)}$, we then have
\begin{equation}
{\color{black}l_i=\min\left((\beta{r}-\frac{r}{\lambda_i})^+,L\right)},
\end{equation}
and $\beta$ can be effectively solved by the bisection  method.
\section*{Appendix D: Proof of Proposition~\ref{pro:CCE-GEBB}}\label{app:CCE=GEBB}
 Let $d_{\textmd{any}}$ be  the waiting time  of an arbitrary proactive delivery mechanism  for $V_i$ with  bandwidth  $b_i$ and cache size  $l_i$, where $e^{\frac{f_B}{r}b_i}l_i\leq L$. The prefix of the video instead of other parts should be cached since the beginning part is always firstly displayed and consumes more bandwidth. The remaining part with length $L-l_i$  is then  periodically broadcasted under bandwidth $b_i$.
 Let $dx$ denote a small portion at an arbitrary length offset $x$ of the $i$-th video, where $l_i\leq x\leq L$.  Note that $dx$ should be successfully buffered within duration $\frac{x}{r}+d_{\textmd{any}}$, and  the corresponding  bandwidth consumption for delivering  partition $dx$ is no less than $\frac{dx}{f_B(\frac{x}{r}+d_{\textmd{any}})}$. Therefore, the  allocated bandwidth is lower bounded by the following expression
\begin{equation}
b_i\geq\int_{l_i}^{L}\frac{dx}{f_B(\frac{x}{r}+d_{\textmd{any}})}=\frac{r}{f_B}\ln\left(\frac{L-l_i}{l_i+rd_{\textmd{any}}}+1\right).
\end{equation}
We then have
\begin{equation}
d_{\textmd{any}}\geq\frac{L-e^{\frac{f_B}{r}b_i}l_i}{r(e^{\frac{f_B}{r}b_i}-1)}=d_{i}^{FA}
\end{equation}
valid for any proactive delivery mechanism. Hence,  CCE-GEBB is optimal to minimize the  waiting time of the $i$-th video with bandwidth $b_i$ and cache size $l_i$.
\section*{Appendix E: Proof of Proposition~\ref{pro:fa_structure}}\label{app:fa_structure}
Firstly, we proof part (a). Given the optimal  bandwidth allocation $\mathbf{b}$,  the cache  problem becomes
\begin{equation}\label{equ:generalproblem}
\begin{split}
    \max_{\mathbf{l}}& ~~\sum\limits_{i=1}^{M}\frac{{p_i}e^{\frac{f_B}{r}b_i}}{r(e^{\frac{f_B}{r}b_i}-1)}l_i \\
    \textup{s.t.}&  ~~\sum\limits_{i=1}^{M}{l_i}\leq C, 0 \leq l_i\leq L/ e^{\frac{f_B}{r}b_i}, \forall i\in\{1,\ldots,M\},\\
\end{split}
\end{equation}
which is a fractional knapsack problem with the optimal greedy solution.   The weight of caching the $i$-th video is $w_i={{p_i}e^{\frac{f_B}{r}b_i}}/{(r(e^{\frac{f_B}{r}b_i}-1))}$. The greedy solution  allocates more cache storage to the video with largest weight until reaching its maximum value, i.e., $l_i=Le^{-\frac{f_B}{r}b_i}$ to provide zero-delay for the $i$-th video. Since the $k$-th video has allocated cache storage and experiences non-zero delay,  the  videos with larger weights should be zero-delay and the ones with smaller weights have no cache allocated due to the greedy property. Then we only need to validate  that a larger popularity represents a larger weight in this case. By contradiction, we assume  that $V_j$ has zero-delay while  $V_i$ with a larger popularity experience  non-zero delay ($i\leq{j}$), the average waiting time will be decreased by simply switching the  cache-bandwidth allocations of these two videos. Therefore, a larger popularity stands for a larger weight and part (a) is proved.

Next, we prove part (b).  For the first $k-1$ videos with zero-delay, the optimal solution  should utilize the  minimum  cache  usage given the total allocated bandwidth  and the minimum bandwidth usage given the total allocated cache.  Firstly, we consider the cache minimization problem given  the total allocated bandwidth $B_s$ for the first $k-1$ videos
\begin{equation}\label{equ:generalproblem}
\begin{split}
    \min_{\mathbf{l}}& ~~\sum\limits_{i=1}^{k-1}Le^{-\frac{f_B}{r}b_i}\\
    \textup{s.t.}&  ~~
    \sum\limits_{i=1}^{k-1}{b_i}=B_s, {b_i}\geq 0, i\in\{1,\ldots,M\},
\end{split}
\end{equation}
    which is a convex problem and can be effectively solved by the Lagrangian method. The optimal solution is that  the bandwidth and the cache are evenly allocated. Same results can also be found for the  total bandwidth minimization problem given the total cache usage for $k-1$ videos. Therefore, for the videos with zero-delay,  bandwidth and   cache are evenly allocated.

\section*{Appendix F: Proof of Lemma~\ref{lemma:fa_optimal}}\label{app:fa_optimal}
Based on \textbf{Proposition \ref{pro:fa_structure}}, the structure of the optimal cache allocation obeys
 \begin{equation}
 l_i=\begin{cases}l_1 \quad &\text{if } 1\leq i \leq k-1,\\
 C\bmod l_1 \quad &\text{if }  i= k,\\
 0 \quad &\text{if } k+1 \leq i \leq M,
 \end{cases}
\end{equation}
where $k=\lfloor{C}/{l_1}\rfloor+1$.

If $l_1<C$, the first $k-1$ videos are zero-delay, yielding $b_i=\frac{r}{f_B}\ln(\frac{L}{l_1})$ for $1\leq i \leq k-1$.  The bandwidth allocation for the remaining $M-k+1$ videos becomes
\begin{align}\label{pro:pro_fa_proof}
    \min_{\mathbf{b}}& ~~\sum\limits_{i=k}^{M}{p_i}d_{i}^{FA}\\
    \textup{s.t.}&  ~~\begin{cases}\sum\limits_{i=k}^{M}{b_i}\leq B-\sum\limits_{j=1}^{k-1}{b_i},\\
     b_i\geq 0, \forall i\in\{k,\ldots,M\}.
    \end{cases}
\end{align}
We have $\frac{\partial^2d_{i}^{FA}}{\partial b_{i}^2}\geq 0$, thus $d_{i}^{FA}$ is convex in $b_i$ and Problem (\ref{pro:pro_fa_proof}) is a convex problem. Considering the  Lagrangian
 \begin{equation}
{\cal{L}}=\sum\limits_{i=k}^{M}{p_i}(\frac{L-l_i}{r(e^{\frac{f_B}{r}b_i}-1)}-\frac{l_i}{r})+\mu(\sum\limits_{i=1}^{M}{b_i}-B),
\end{equation}
where $\mu$ is the Lagrange multiplier. The KKT condition for the optimality of a bandwidth allocation for the remaining $M-k+1$ videos becomes
\begin{align}
\frac{\partial\cal{L}}{\partial{b_i}}=-\frac{f_Be^{\frac{f_B}{r}b_i}(L-l_i)}{r^2(e^{\frac{f_B}{r}b_i}+1)^2}+\mu\begin{cases}=0 \quad \text{if }  b_i>0,\\
\geq 0 \quad \text{if } b_i=0.
\end{cases}
\end{align}
Let $\beta={f_B}/({{\mu}r^2})$, we then have, for $ i=k,\ldots,M$,
\begin{equation}\label{equ:pro_fa_mu}
\small{b_i\!=\!\frac{r}{f_B}\ln(\frac{2+{p_i\beta(L-l_{i})}\!+\!\sqrt{(p_i\beta(L-l_{i}))^2\!+\!4{p_i\beta(L\!-\!l_{i})}}}{2})}.
\end{equation}

For videos with no cache  allocated, more bandwidth is assigned to videos with larger popularity, i.e.,  $b_k\geq\ldots\geq b_M$.

When $l_1=C\leq l$, the whole cache size is allocated to the first video, and  $b_1$ also obeys   (\ref{equ:pro_fa_mu}).

Therefore, the optimization can be found by the one dimension search of $l_1$, where  $l_1\in[L/M, \min(L,C)]$.
\section*{Appendix G: Proof of {\color{black} Proposition~\ref{theorem:FD_cache}}}\label{app:pro1}
By using Cauchy-Schwarz Inequality, we have
{\small{$\left(\sum_{i=1}^{M}\frac{p_i\left(L-l_i\right)}{f_Bb_i}\right)\left(\sum_{i=1}^{M}b_i\right)\geq \left(\sum_{i=1}^{M}\sqrt{\frac{p_j\left(L-{l_j}\right)}{f_B}}\right)^2$}}, yielding
\begin{equation}
\sum\limits_{i=1}^{M}\frac{p_i\left(L-l_i\right)}{f_Bb_i}\geq \frac{\left(\sum\limits_{i=1}^{M}\sqrt{p_j\left(L-{l_j}\right)}\right)^2}{f_BB},
\end{equation}
where the equation is achieved when {\small{$b_i={B\sqrt{p_i(L-{l_i})}}/{\sum_{j=1}^{M}\sqrt{p_j(L-{l_j})}}$}}.  The  problem requires minimizing {\small{$\sum_{i=1}^{M}\sqrt{p_j\left(L-{l_j}\right)}$}}, which can be effectively solved by the following proposition.
\begin{proposition}
The optimal solution has the following two properties: a) ${l_1\geq l_2 \ldots\geq l_M}$. b) if $0<l_j<L$, we have $l_i=L$ for $i=1,2,\ldots,j-1$.
\end{proposition}
\begin{IEEEproof}
Firstly, we assume by contradiction that there exists $l_i<l_j$ for $i<j$, and {\small{$\sum_{i=1}^{M}\sqrt{p_j\left(L-{l_j}\right)}$}} will be reduced by simply switching the allocations for $V_i$ and $V_j$. Thus part (a) is proved.

Secondly, we assume by contradiction that there exists $l_i<L$ for $i<j$. By shifting  a cache storage size $\Delta$ of $V_j$ to $V_i$, where $\Delta=\min\{l_j,L-l_i\}$, we only need to prove {\small{$\sqrt{p_i\left(L-(l_i+\Delta)\right)}+\sqrt{p_j\left(L-(l_j-\Delta)\right)}<\sqrt{p_i\left(L-l_i\right)}+\sqrt{p_j\left(L-l_j\right)}$}}.

Denote {\small{$f(x)=\sqrt{p_i\left(L-(l_i+\Delta)+{x}\right)}-\sqrt{p_j\left(L-l_j+{x}\right)}$}}, where $p_i>p_j$, $l_i<l_j$ and $\Delta>0$.
We then have $f'(x)>0,$
thus $f(x)$  increases with $x$ and we have $f(0)<f(\Delta)$, yielding
$\sqrt{p_i\left(L-(l_i+\Delta)\right)}-\sqrt{p_j\left(L-l_j\right)}<\sqrt{p_i\left(L-l_i\right)}-\sqrt{p_j\left(L-(l_j-\Delta)\right)}$. Thus the optimal cache allocation in this case is to cache the most popular videos only.
\end{IEEEproof}
\bibliographystyle{IEEEtran}
\bibliography{VOD_paper}
\end{document}